\documentclass{ws-procs10x7}

\begin{document}

\title{Prompt Photon Production at HERA and LEP}

\author{Thomas Kluge (on behalf of the H1, OPAL and ZEUS Collaborations)}

\address{DESY, Notkestr. 85, 22607 Hamburg, Germany\\E-mail: thomas.kluge@desy.de}

\twocolumn[\maketitle\abstract{
Results on isolated prompt photon production are presented.
The measurements were performed at HERA in deep inelastic $ep$ scattering and photoproduction,
as well as at LEP in photon photon collisions.
Differential cross sections are shown for inclusive prompt photons and those accompanied by a jet.
The results are compared to predictions of perturbative QCD calculations in next to leading order and
 to predictions of the event generators PYTHIA and HERWIG. 
}]

\section{Introduction}
Prompt photons in the final state of high energy collisions allow for a detailed study of perturbative QCD
(pQCD)
 and the  hadronic structure of the incoming particles.
The term ``prompt'' refers to photons which are radiated directly from the partons of the hard interaction,
 instead of stemming from the decay of hadrons or from QED radiation from electrons.

In contrast to jets, photons are not affected by hadronisation, resulting in a more direct correspondence
 to the underlying partonic event structure.
Moreover, the experimental uncertainties connected to the energy determination of an electromagnetic shower
initiated by a photon are smaller compared to the measurement of jets of hadrons.
However, the cross section for isolated prompt photon production is small and the identification of photons in the detector
 is elaborate.

In the following, results from the HERA experiments H1 and ZEUS are presented, together with findings from OPAL at LEP.
The OPAL Collaboration studies isolated prompt photons in the collisions of quasi real photons at $e^+e^-$ centre-of-mass energies
  between $183\,\mathrm{GeV}$ and $209\,\mathrm{GeV}$.
The HERA experiments exploit photoproduction and deep inelastic scattering at an $ep$ centre-of-mass energy of $319\,\mathrm{GeV}$.

Measurements are presented for the inclusive production of prompt photons as well as for the case where a jet in addition is required.
The inclusion of jet observables allows to distinguish between processes where incoming photons
 interact point like  (``direct'')  or exhibit a hadronic structure (``resolved'').

\section{Photon Identification}
Photons are identified by a three step procedure, which is similar for the different analyses presented here.
First a basic photon selection is performed, where an electromagnetic cluster is required in the central calorimeter.
In order to suppress electrons, a veto on tracks associated with this cluster is applied. 

At this stage, a large fraction of the event sample is still made up of background  $\pi\to\gamma\gamma$ and $\eta\to\gamma\gamma$,
 especially at high $E_t$ of the photon candidate, where the opening angle of the photons is small and
 the pair of particles cannot be resolved.

Most hadrons are produced together with a jet, hence in a second step the photon candidate is
 required to be isolated from other particles in the $\eta-\phi$ plane\footnote{The pseudo-rapidity is defined as $\eta=-\log(\tan \theta/2)$ with $\theta$ the polar angle w.r.t to the z-axis. In the H1 and ZEUS coordinate systems the z-axis points in the direction of the proton beam, while in the OPAL coordinate system the z-axis points in the direction of the electron beam}.
This procedure also introduces a sensitivity to energy deposits on top of the hard process, e.g.\ to multiple interactions (m.i.).

As the final step, a shower shape analysis is performed on the candidate cluster, exploiting the fact
 that single photons are more compact than photon pairs from decays.
This makes use of the lateral and longitudinal resolution of the electromagnetic calorimeter.
With help of estimators constructed from the shower shape parameters, the ratio of the signal over the background
 in the data is fitted in every result bin.
By this the result is independent from the background rate in the Monte Carlo used for the fit.

\section{Results}
\begin{figure}
\epsfxsize120pt
\includegraphics[width=0.23\textwidth]{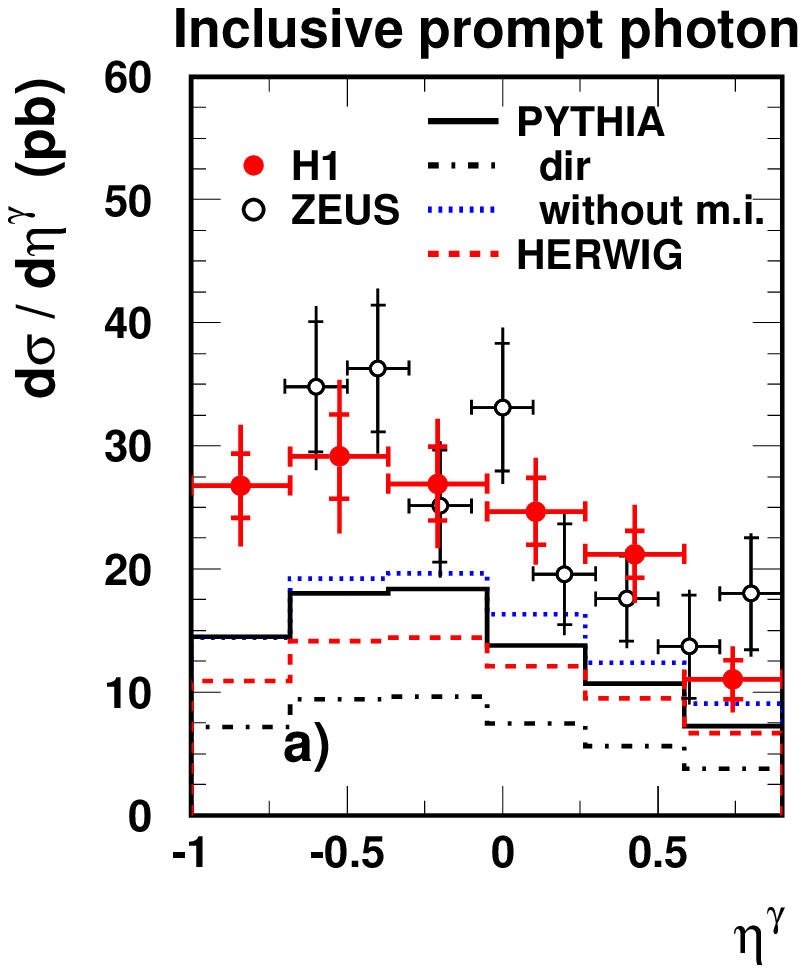}
\includegraphics[width=0.23\textwidth]{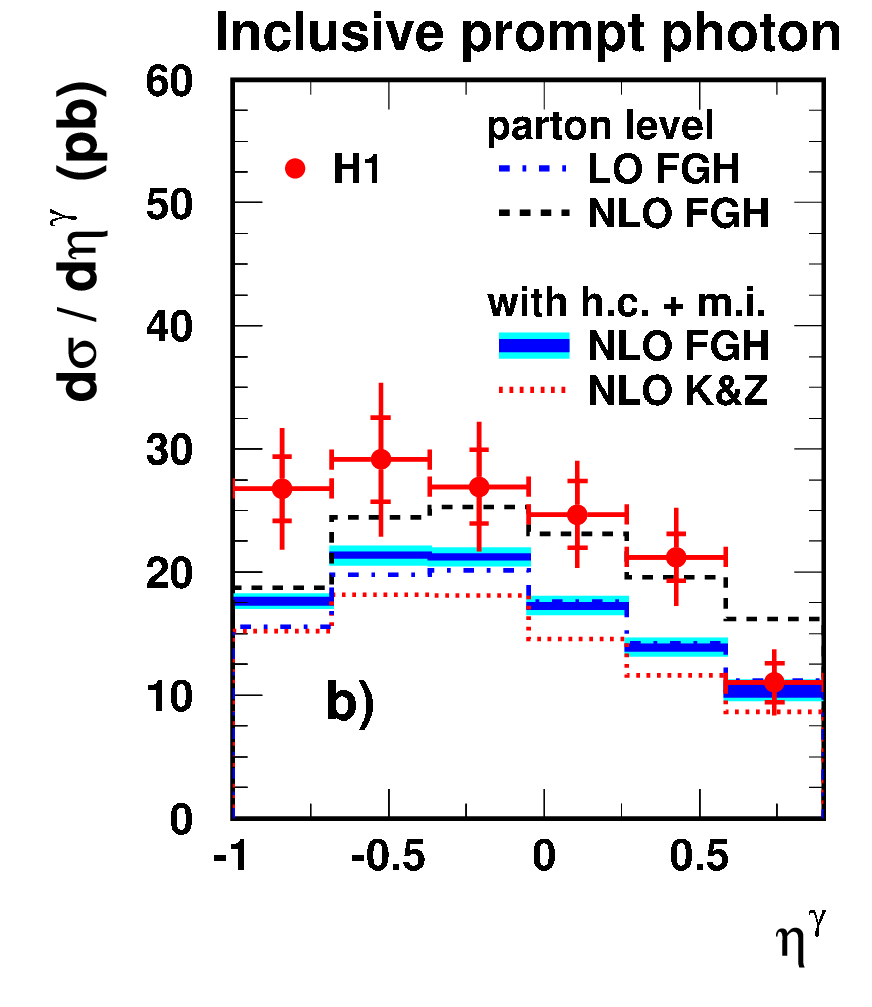}
\caption{Inclusive prompt photon differential cross section for $-1<\eta^\gamma<0.9$ in photoproduction,
 $\sqrt{s}=319\,\mathrm{GeV}$ and $Q^2<10\,\mathrm{GeV}^2$ from the H1 and ZEUS Collaborations.
a) Comparison to HERWIG and PYTHIA including multiple interactions, only the direct contribution (dir) and the prediction without
 multiple interactions (m.i.).
 b) Comparison with NLO pQCD calculations (K\&Z
and FGH,
see text),
 corrected for hadronisation and multiple interaction (h.c. + m.i.) effects.
}
\label{fig:1}
\end{figure}

Fig.~\ref{fig:1} shows the differential inclusive prompt photon cross section $\mathrm{d}\sigma/\mathrm{d}\eta^\gamma$ as a function
 of the pseudorapidity of the photon.
The data were obtained in $ep$ photoproduction ($Q^2<1\,\mathrm{GeV}^2$ and $142<W<266\,\mathrm{GeV}$).
Measurements by the H1\cite{Aktas:2004uv} and ZEUS\cite{Breitweg:1999su} Collaborations are found
 to be consistent within errors.
A comparison with predictions by PYTHIA\cite{Sjostrand:2000wi} and HERWIG\cite{Corcella:1999qn} finds the shape of the distribution 
 reasonably described, but the normalisation $40\%-50\%$ too low.
The predictions by PYTHIA are shown with and without multiple interactions.
The latter results in a decrease of the cross section, because additional energy may get into the isolation cone.

Also shown is a comparison of the H1 data with next to leading order (NLO) pQCD calculations by Fontannaz, Guillet and Heinrich (FGH\cite{Fontannaz:2001ek}) 
 and Krawczyk and Zembrzuski (K\&Z\cite{Krawczyk:2001tz}).
The calculations use the photon and proton parton density functions AFG\cite{Aurenche:1994in} and MRST2\cite{Martin:1999ww},
 respectively, and BFG\cite{Bourhis:1997yu} fragmentation functions.
On parton level the calculations provide a reasonable description of the data.
However, after corrections determined with PYTHIA for hadronisation and multiple interactions have been applied,
 the normalisation is found to be $20\%-40\%$ below the data.

\begin{figure}
\epsfxsize120pt
\includegraphics[width=0.23\textwidth]{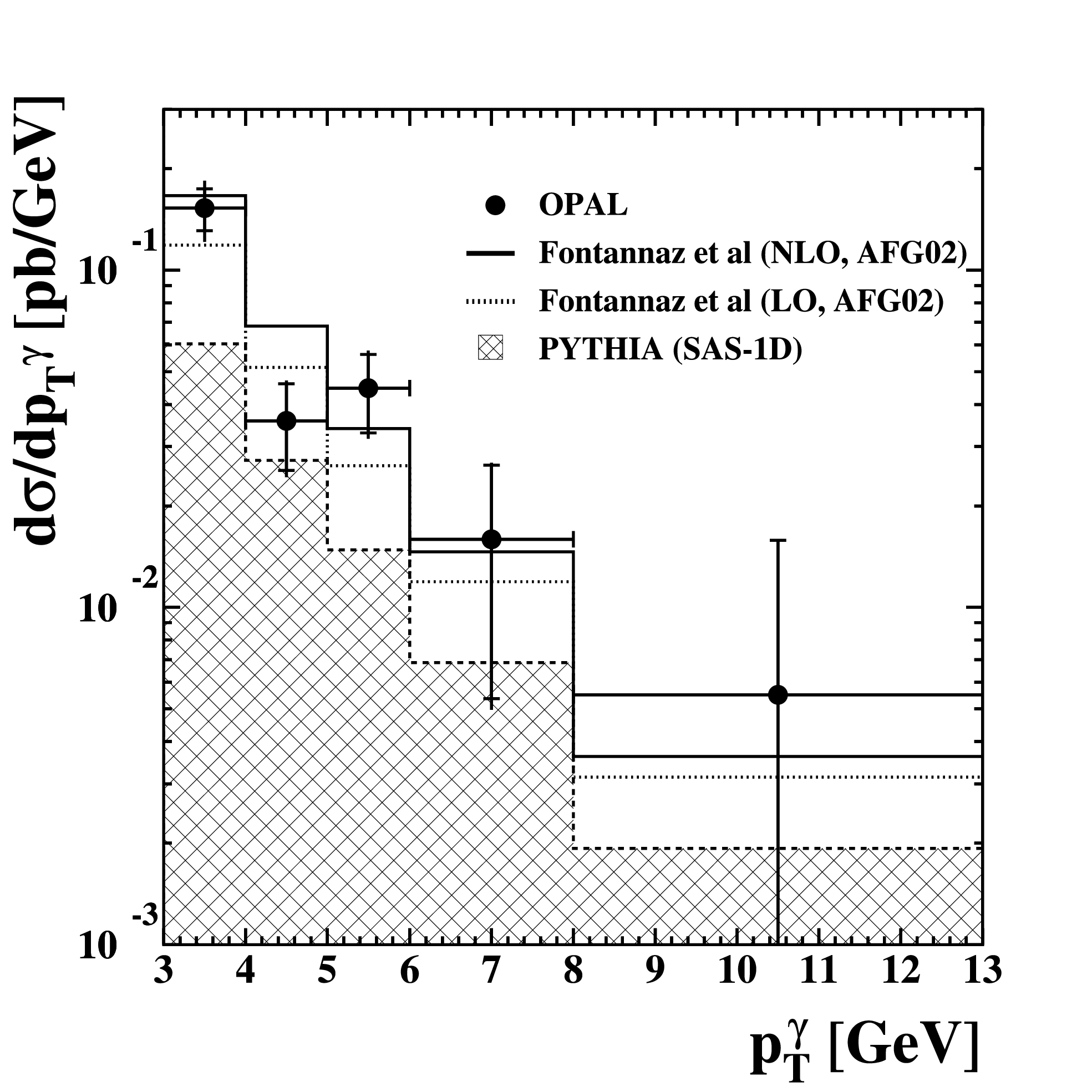}
\includegraphics[width=0.23\textwidth]{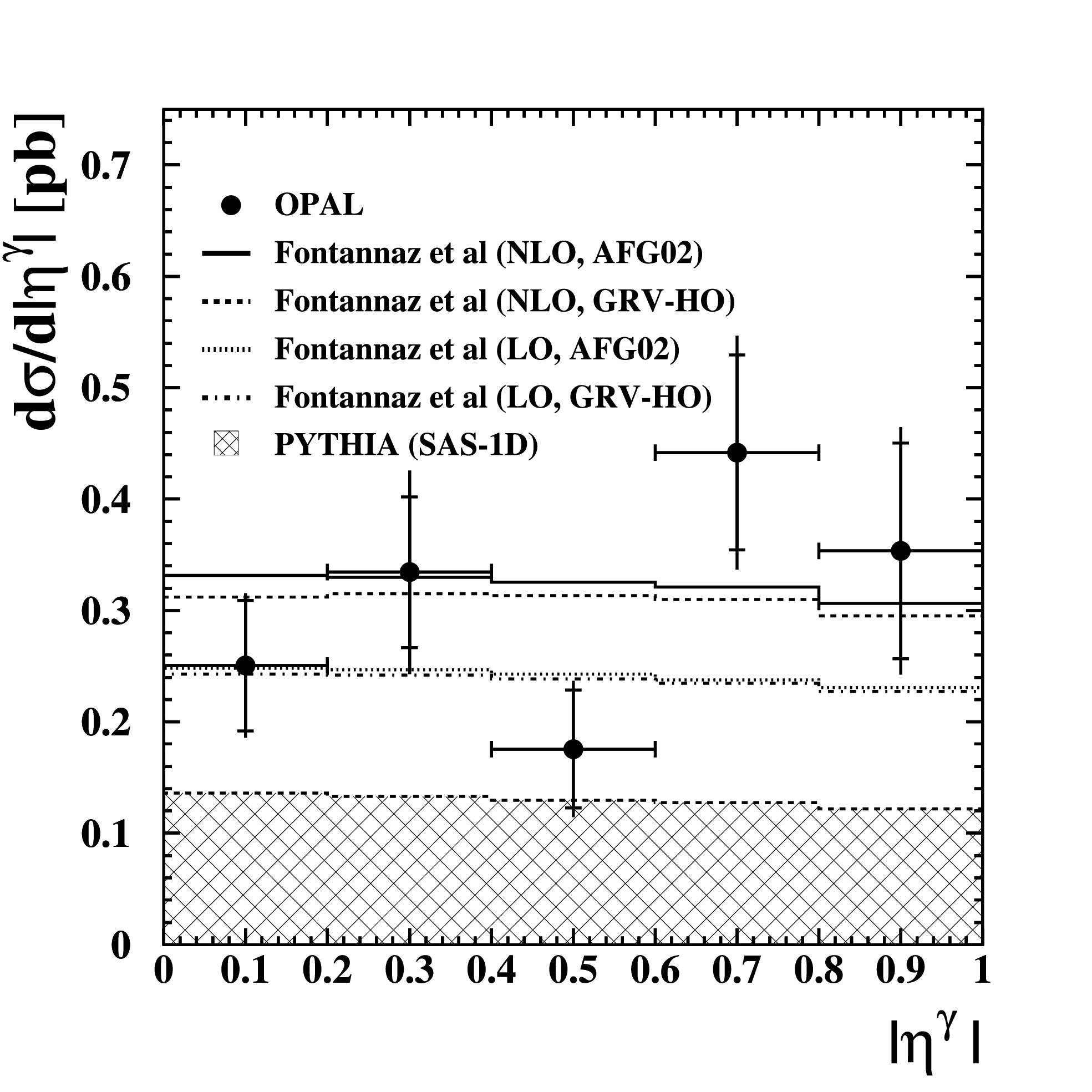}
\caption{Inclusive prompt photon differential cross section for $|\eta^\gamma|<1$ and $p^\gamma_T>3\,\mathrm{GeV}$ 
in photon photon collisions, $\sqrt{s_{ee}}=196\,\mathrm{GeV}$ and $Q^2<10\,\mathrm{GeV}^2$
 from the OPAL Collaboration. The data are compared to PYTHIA and a NLO pQCD calculation, using two parametrisations
 of the photon pdfs (AFG02 and GRV-HO).}
\label{fig:2}
\end{figure}

Fig.~\ref{fig:2} depicts the inclusive prompt photon cross section measured by OPAL\cite{Abbiendi:2003kf}, differential 
 in the transverse momentum and pseudorapidity of the photon.
The data measured in photon photon collision ( $\sqrt{s_{ee}}=196\,\mathrm{GeV}$ and $Q^2<10\,\mathrm{GeV}^2$)
 are compared to PYTHIA and NLO pQCD calculations performed by Fontannaz et al\cite{Fontannaz:2001kf}.
PYTHIA gives a good description of the shape, but the normalisation of the distributions is low by $~50\%$.
On the other hand, the NLO calculation on parton level describes the data rather well.
Two sets of parton density functions of the photon were used (AFG02\cite{Aurenche:1994in} and GRV-HO\cite{Gluck:1991jc}), 
whereby the resulting change in the predictions is small compared to the uncertainty of the data.

\begin{figure}
\epsfxsize120pt
\includegraphics[width=0.23\textwidth]{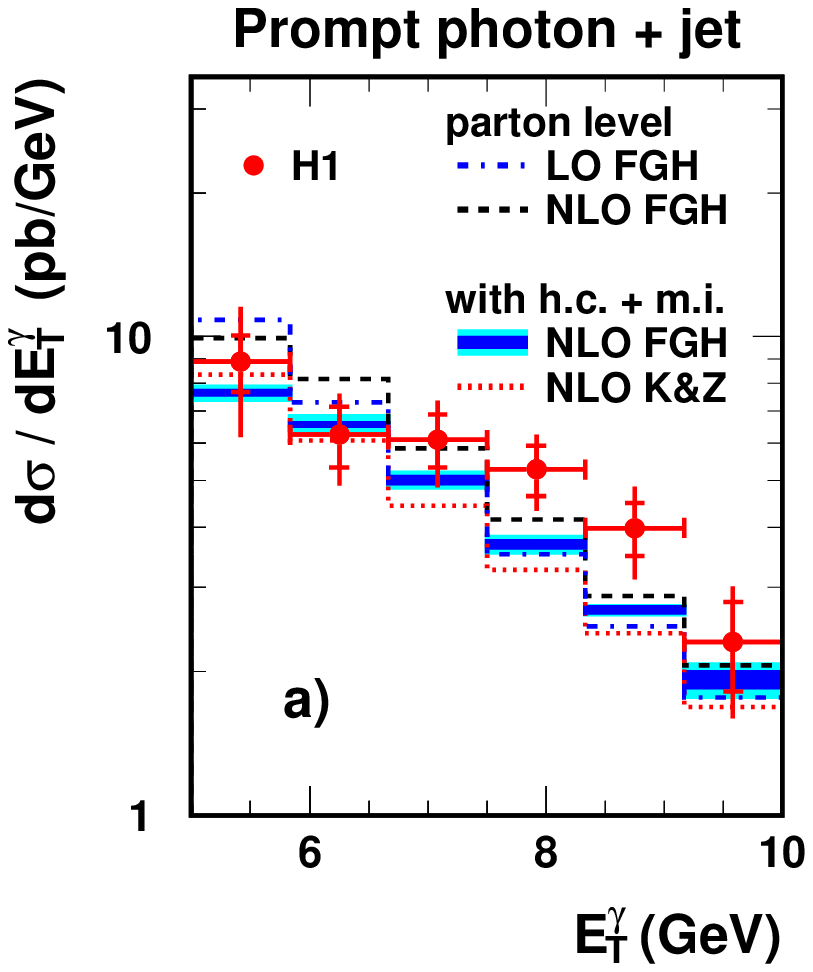}
\includegraphics[width=0.23\textwidth]{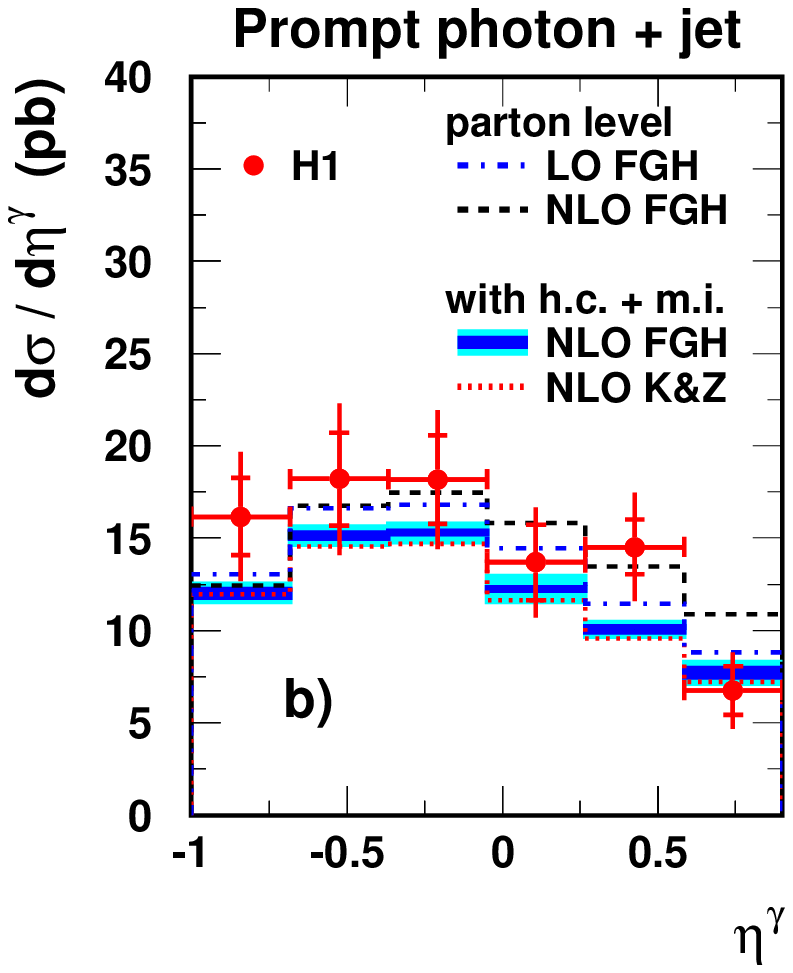}
\caption{Prompt photon cross sections in photoproduction with an additional jet requirement
 ($E_{T}^{jet}>4.5\,\mathrm{GeV}$ and $-1<\eta^{jet}<2.3$).
Data from the H1 Collaboration are compared with NLO pQCD calculations.}
\label{fig:3}
\end{figure}

Fig.~\ref{fig:3} shows cross sections measured by the H1 Collaboration in photoproduction, where an additional jet ($E_t>4.5\,\mathrm{GeV}$ and 
$-1<\eta^{jet}<2.3$) is required.
The  differential cross section as a function of the transverse momentum and the pseudorapidity of the photon are
  compared to the pQCD calculations.
In contrast to the inclusive case, both NLO calculations\cite{Fontannaz:2001nq,Zembrzuski:2003nu} are consistent with the data in most bins,
 and also the NLO/LO corrections are reduced. 
The hadronic and m.i.\ corrections improve the description of the data only in some regions.

\begin{figure}
\epsfxsize120pt
\includegraphics[width=0.23\textwidth]{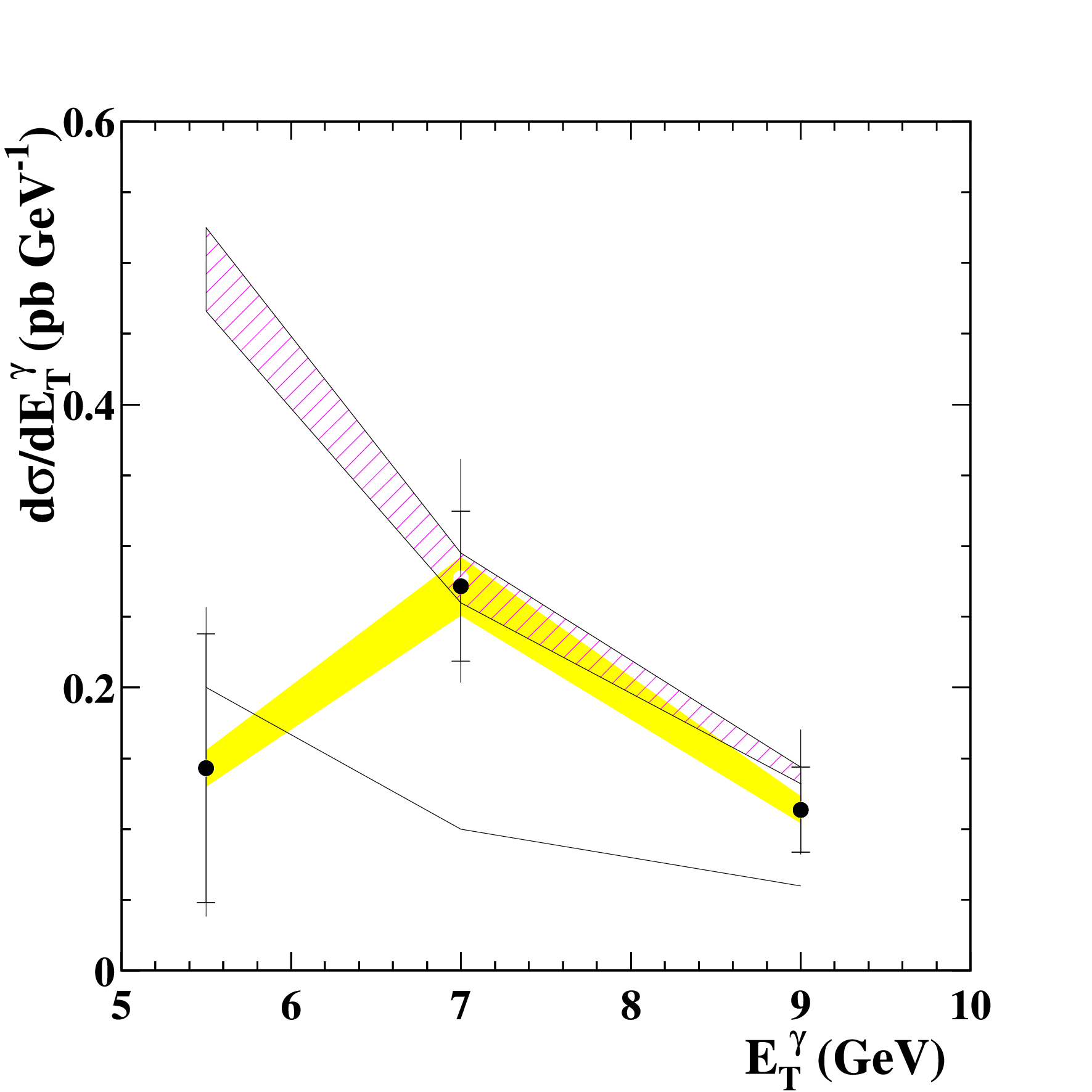}
\includegraphics[width=0.23\textwidth]{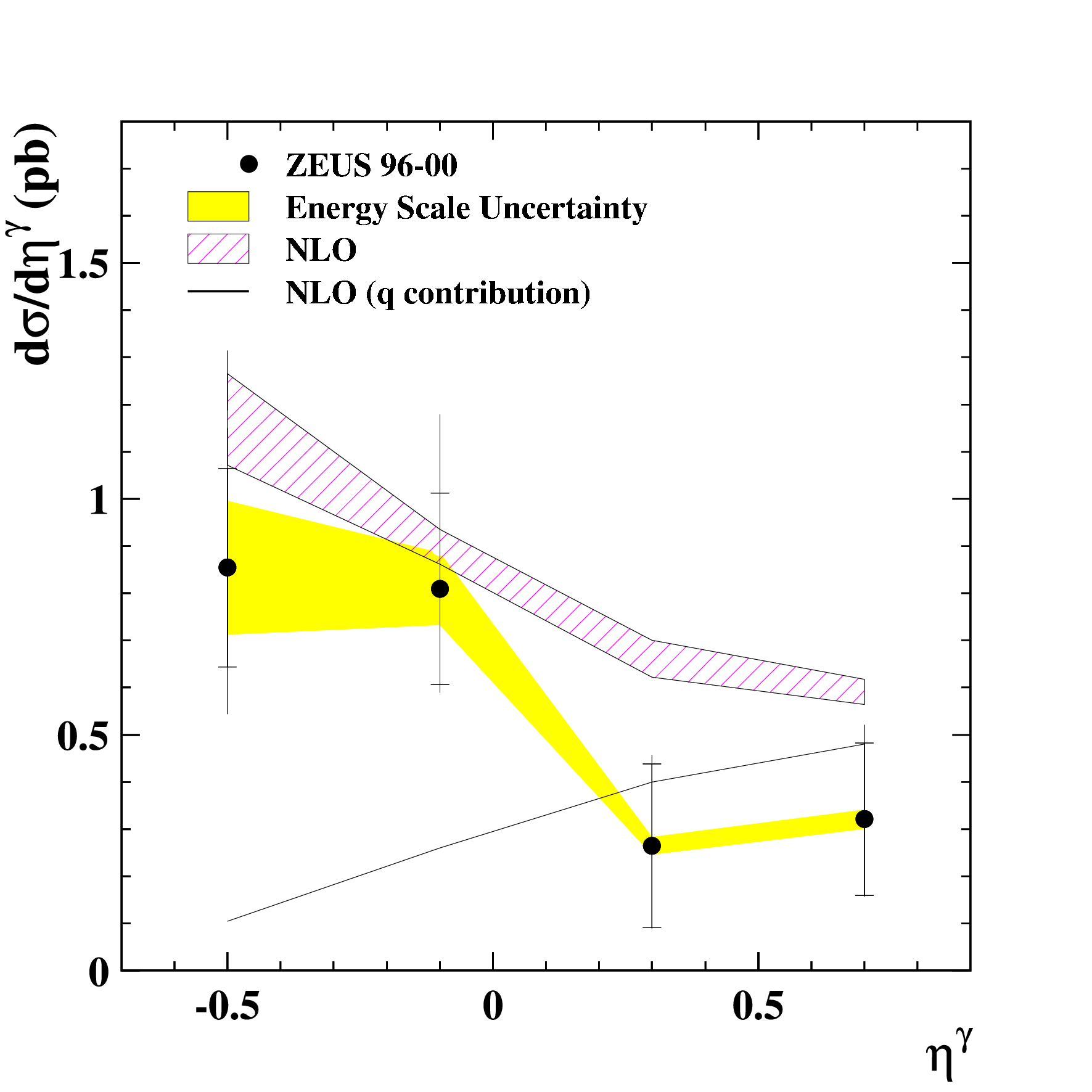}
\caption{Prompt photon cross sections in DIS ($Q^2>35\,\mathrm{GeV}^2$ ) 
with an additional jet requirement ($E_{T}^{jet}>6,\mathrm{GeV}$ and $-1.5<\eta^{jet}<1.8$)
Data from the ZEUS Collaboration are compared with a NLO pQCD calculation.}
\label{fig:4}
\end{figure}

The ZEUS Collaboration\cite{Chekanov:2004wr} has measured the inclusive prompt photon in DIS ($Q^2>35\,\mathrm{GeV}^2$), shown in Fig.~\ref{fig:4}.
In addition to the prompt photon a jet is required ($E_{T}^{jet}>6\,\mathrm{GeV}$ and $-1.5<\eta^{jet}<1.8$).
A pQCD calculation\cite{Gehrmann-DeRidder:2000ce} to order $\mathcal{O}(\alpha^3\alpha_s^1)$ on parton level 
describes the normalisation of the data rather well. 
However, the description of the shape is poor at low $E_T$ and in the more forward (proton beam) direction.

In order to distinguish direct and resolved processes, variables $x_\gamma^{LO}$ (H1) and $x_{LL}^-$ (OPAL) are introduced:
\begin{displaymath}
x_\gamma^{LO}=\frac{E_T^{\gamma}(e^{-\eta^{\mathrm{jet}}}+e^{-\eta^{\gamma}})}{2yE_e},
\end{displaymath}
\begin{displaymath}
x_{LL}^{-}=\frac{p_T^{\gamma}(e^{-\eta^{\mathrm{jet}}}+e^{-\eta^{\gamma}})}{y^-\sqrt{s_{ee}}},
\end{displaymath}

where $p_T$, $E_t$ and $\eta$ denote the transverse momentum, transverse energy and the pseudorapidity of the
 jet and prompt photon, respectively.
$y$ is the inelasticity , $E_e$ the electron beam energy and $y^-=E_\gamma/E_e$ the fractional energy of 
the quasi-real initial photon oriented towards the negative $z$ axis.
Low values of $x$ denote resolved photon processes, whereas values around one correspond to direct photon interactions.

\begin{figure}
\epsfxsize30pc
\includegraphics[width=0.23\textwidth]{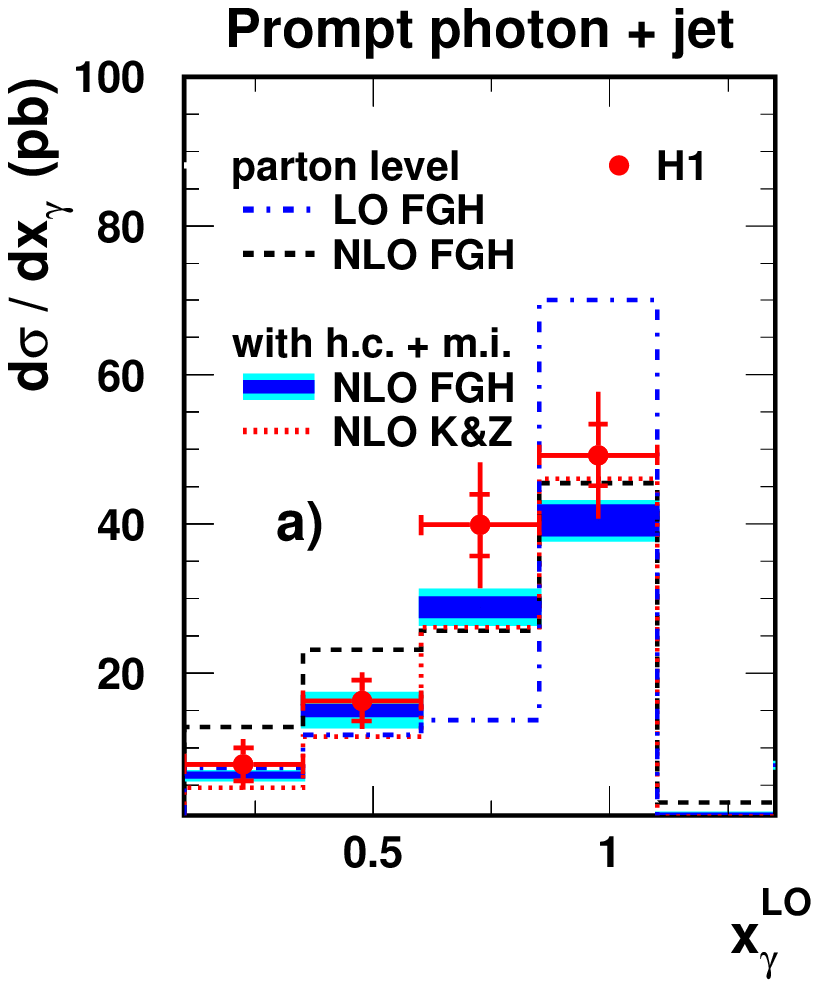}
\includegraphics[width=0.24\textwidth]{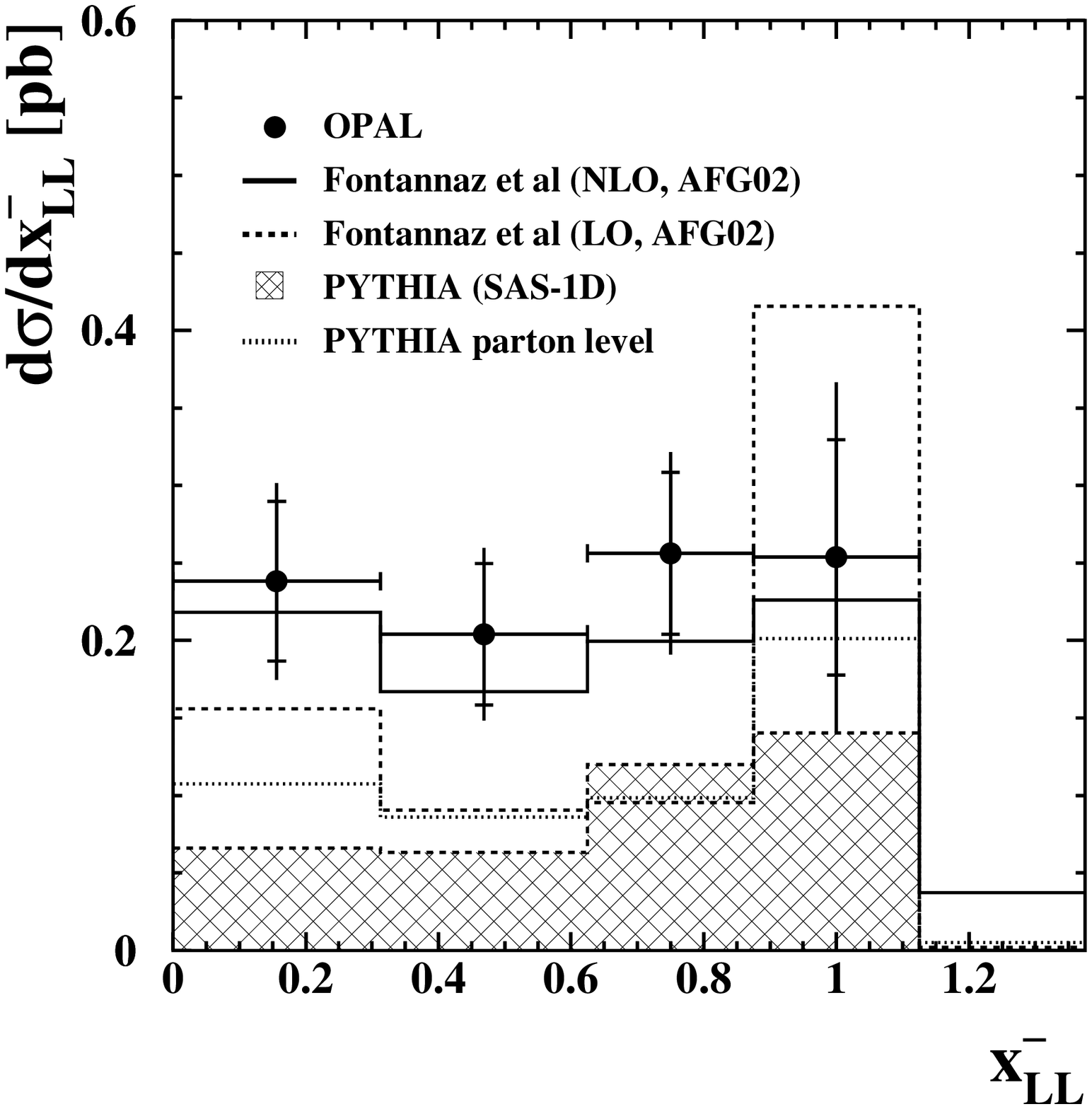}
\caption{Prompt photon cross section with an additional jet requirement in $ep$ photoproduction (left hand side) 
and photon photon collisions (right hand side), differential in $x_\gamma^{LO}$ resp. $x_{LL}^-$, which approximates
 the fractional part of the incoming photon energy, taking part in the interaction.}
\label{fig:5}
\end{figure}

Fig.~\ref{fig:5} shows results from the H1 and OPAL collaborations.
The differential cross section rises towards large values of $x$ for the $ep$  photoproduction case,
in contrast to the photon photon collision measurement, which is compatible with a flat distribution.
In both cases the NLO pQCD calculations are consistent with the data.
The H1 measurement shows that multiple interactions become more important at low $x$, corresponding to
 the regime of resolved photon processes.

\section{Conclusion}
The production of isolated prompt photons has been studied in photoproduction, deep inelastic scattering and 
 photon photon collisions.
In general the predictions of the Monte Carlo event generators PYTHIA and HERWIG undershoot the data, while the shape is rather well described.
NLO pQCD calculations on parton level are in reasonable agreement with the data.
However, if multiple interactions and hadronisation effects are taken into account, the NLO calculations somewhat undershoot the data.

\end{document}